\documentclass[conference]{IEEEtran}
\usepackage{times} 
\usepackage{helvet}  
\usepackage{courier}  
\usepackage[hyphens]{url}  
\usepackage{graphicx}
\usepackage{algorithm}
\usepackage{times}
\usepackage{helvet}
\usepackage{courier}
\usepackage{amssymb}
\usepackage{amsmath}
\usepackage{amsmath, algorithm, algpseudocode}
\usepackage{enumitem}
\usepackage{listings}
\usepackage{xcolor}
\usepackage{booktabs}
\usepackage{multirow}
\usepackage{caption} 
 \usepackage{listings}
\usepackage{xcolor}
\usepackage{tcolorbox}
\usepackage{tcolorbox}
\usepackage[numbers,sort&compress]{natbib} 
\usepackage{caption} 
\tcbuselibrary{listings, skins}

\tcbuselibrary{listings, breakable}

\lstdefinelanguage{json}{
  morestring=[b]",%
  morecomment=[l]{//},
  morekeywords={true,false,null},
  stringstyle=\color{red},
  keywordstyle=\color{blue},
  commentstyle=\color{gray},
  basicstyle=\ttfamily,
  backgroundcolor=\color{gray!15},
  frame = single,
  rulecolor=\color{black}
}
\usepackage{newfloat}
\usepackage{listings}
\DeclareCaptionStyle{ruled}{labelfont=normalfont,labelsep=colon,strut=off} 
\floatstyle{ruled}
\newfloat{listing}{tb}{lst}{}
\floatname{listing}{Listing}
\begin{document}
\title{XGen-Q: An Explainable Domain-Adaptive LLM Framework with Retrieval-Augmented Generation for Software Security}
\author{
Hamed Jelodar, Mohammad Meymani, Roozbeh Razavi-Far, Ali Ghorbani\\
\textit{Canadian Institute for Cybersecurity} \\
\textit{Faculty of Computer Science} \\
\textit{University of New Brunswick} \\
Fredericton, Canada \\
\{h.jelodar, mohammad.meymani79, roozbeh.razavi-far, ghorbani\}@unb.ca
}

\maketitle

\begin{abstract}
Generative AI and large language models (LLMs) have shown strong capabilities in code understanding, but their use in cybersecurity, particularly for malware detection and analysis, remains limited. Existing detection systems often fail to generalize to obfuscated or previously unseen threats, underscoring the need for more adaptable and explainable models. To address this challenge, we introduce XGen-Q, a domain-adapted LLM built on the Qwen-Coder architecture and pretrained on a large-scale corpus of over one million malware samples, spanning both source and assembly code. XGen-Q uses a multi-stage prompt strategy combined with retrieval-augmented generation (RAG) to deliver reliable malware identification and detailed forensic reporting, even in the presence of complex code obfuscation. To further enhance generalization, we design a training pipeline that systematically exposes the model to diverse obfuscation patterns. Experimental results show that XGen-Q achieves significantly lower perplexity than competitive baselines and exhibits strong performance on novel malware samples, demonstrating the promise of LLM-based approaches for interpretable and robust malware analysis.
\end{abstract}
\section{Introduction}

As software systems become more complex and interconnected, the attack surface for security vulnerabilities continues to grow. Modern threat actors increasingly exploit these vulnerabilities using sophisticated malware capable of evading traditional detection mechanisms \cite{bohme2025software,rahimi2025comprehensive}. While static and dynamic code analysis techniques remain essential, they often struggle to detect obfuscated, polymorphic, or previously unseen threats at scale. This has led to a growing interest in AI-driven solutions that offer adaptability, contextual reasoning, and semantic understanding of software behavior \cite{mohamed2025artificial,aldasoro2025generative,rahman2025cyber}.
\begin{figure}[h]
    \centering
    \includegraphics[width=\linewidth]{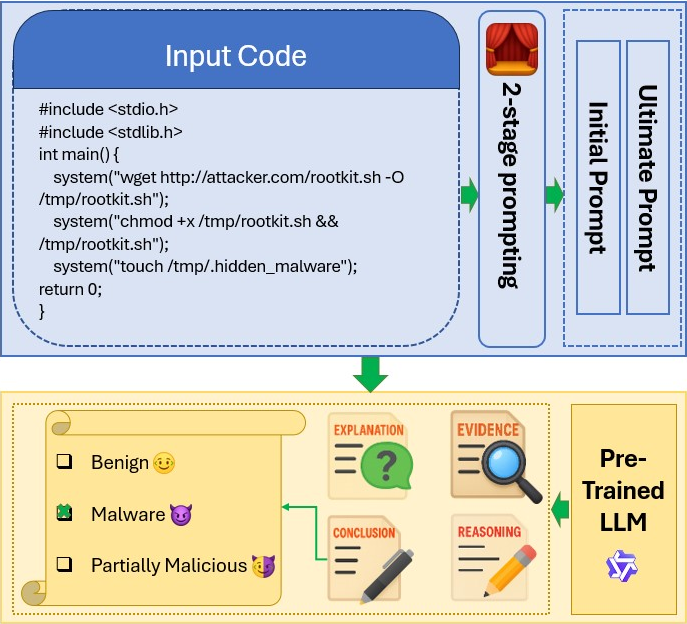}
    \caption{This figure illustrates a simple interaction between an input code and our proposed framework, where the system has labeled the input as malware.}
    \label{fig:XGen-Q}
\end{figure}
Recent advances in LLMs have demonstrated exceptional performance in code understanding, generation, and reasoning tasks \cite{xu2024large,tian2025exploring, kasri2025vulnerability}. Also, there are some work related to malware code analysis \cite{hossain2024malicious,al2024exploring,zhou2025srdc}. For instance, in
\cite{qian2025lamd}, the authors focused on overcoming the challenges of applying LLMs to Android malware detection, such as large codebases and complex app structures.  In other work \cite{yu2024maltracker}, the authors focused on improving malware detection by using fine-grained code features and expanding the dataset with LLM-translated malicious functions from other languages.\\

However, their direct application to cybersecurity remains limited, especially in areas requiring behavioral pattern recognition, threat explanation, and integration with operational workflows. These limitations originate from challenges in dataset availability, model interpretability, and the lack of end-to-end frameworks that map LLM outputs with real-time cybersecurity workflows \cite{xu2024large}. \\

Although general-purpose LLMs have shown promises on malware detection, malware classification, and malware analysis, they are trained predominantly on large-scale but generic programming corpora, often lack the specialized knowledge needed to assess malicious intent or accurately interpret attacker's strategies \cite{kasri2025vulnerability,hasanov2024application,zhang2025llms}.\\ 

In this work, we propose XGen-Q (Explainable Generation-Driven Qwen Model for Malware Behavior Analysis), a domain-adaptive, retrieval-augmented generative AI framework designed to strengthen prevention, mitigation, and preservation strategies in software security. The primary contributions and innovations of this work are as follows:

\begin{itemize}
    \item We introduce XGen-Q, a systematic large language model framework for malware behavior analysis, trained on a diverse corpus of real-world malware samples in both assembly and source code. This domain-specific pretraining allows the model to capture low-level behavioral patterns that general-purpose LLMs often fail to recognize.
    
    \item We design a retrieval-augmented generation (RAG) mechanism to dynamically incorporate external cybersecurity knowledge during inference. This improves the model’s contextual awareness and adaptability to emerging threats through behavior keyword extraction and prompt augmentation.

    \item We develop a two-stage prompt architecture that separates structured forensic reporting from final behavior classification. This promotes interpretability, transparency, and operational flexibility by providing both human-readable reports and system-ready verdicts (malware, benign, partially malicious).
\end{itemize}
\footnote{The code and LLM framework are open source and available here: https://huggingface.co/JeloH/xGenq-qwen2.5-coder-1.5b-instruct-OKI}
The rest of the paper is organized as follows: In Section related-works, we review the existing literature on LLMs for code analysis and malware detection. In Section methodology, the details of XGen‑Q framework, including data preparation, domain‑specific pre‑training, model architecture, and multi‑stage prompt design are explained. In Section experiments and implementation, we present our experimental setup, evaluation metrics, and implementation details and discusses XGen‑Q’s performance on malware behavior classification and forensic analysis tasks. In Section limitations and future works, we examine the current limitations of our approach and outlines directions for future improvement. Finally, in Section conclusion, we conclude the paper and highlights promising avenues for further research.

\section{Related Works}
\label{sec:related-works}
The intersection of LLMs and software security analysis has recently attracted significant attention. Several works explore leveraging pre-trained LLMs for detecting malicious code, understanding malware behavior, and assisting in reverse engineering tasks.
\begin{figure*}[h]
\centering
\includegraphics[width=\linewidth]{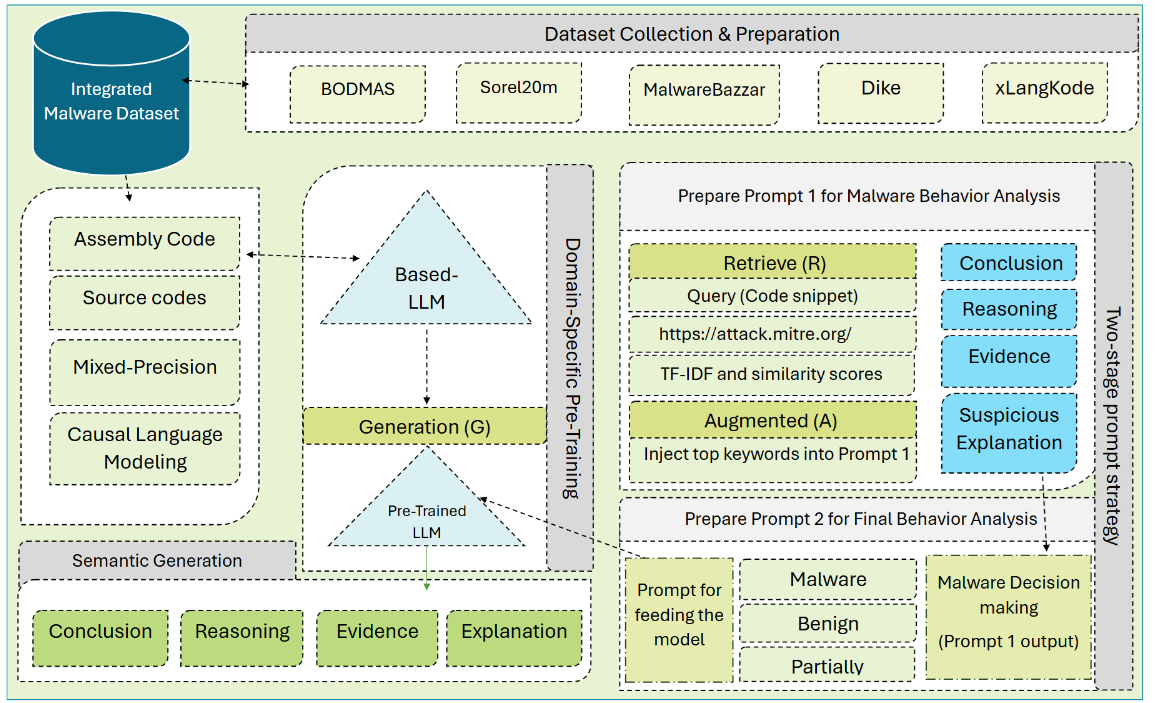}
\caption{The general diagram of the proposed framework.}
\label{fig:diagram-of-framework}
\end{figure*}

\textbf{LLMs for Malware Detection and Analysis:}  
Recent studies \cite{hossain2024malicious,feng2025llm,qian2025lamd,zhou2025srdc} have demonstrated the effectiveness of adapting general-purpose LLMs to malware detection by pre-training on domain-specific datasets. These approaches typically focus on improving detection accuracy and interpretability by incorporating behavioral patterns extracted from static and dynamic analysis. However, there existing challenges in handling obfuscated and polymorphic malware, which demand more nuanced contextual understanding.

\textbf{Domain-Adaptive Pretraining and Retrieval-Augmented Generation:}  
To enhance LLM performance in specialized domains, such as cybersecurity, domain-adaptive pretraining has been proposed and is used by several works \cite{lee2025trag,long2023adapt,leemann2024auto,xu2024simrag}. This involves continued training on security-specific corpora, enabling better modeling of code semantics and malicious patterns. RAG techniques have further improved inference by integrating external knowledge dynamically \cite{lewis2020retrieval}. These strategies help models stay up-to-date with emerging threats and mitigate knowledge cutoff issues inherent to static pretraining. 

\textbf{Perplexity as a Performance Metric in Code Modeling:}  
Perplexity has been widely used as an intrinsic metric to evaluate the predictive capability of language models on code and natural language tasks. Lower perplexity generally correlates with better token prediction and semantic understanding \cite{brown2020language}. In the context of malware and code analysis, perplexity helps quantify how well models grasp complex, obfuscated code patterns, making it a valuable benchmark for comparing competing models. Works such as \cite{xu2024detecting,cooper2024perplexed,xu2024investigating,yusof2024code} explored this area. To the best of our knowledge, this is the first work to unify domain-specific pretraining, retrieval-augmented generation, and multi-stage prompt engineering for malware analysis with LLMs. In addition to achieving state-of-the-art performance on perplexity benchmarks, our framework enhances interpretability and resilience against evolving threat landscapes. This contribution represents a meaningful stride toward integrating the progress of modern language modeling with the practical demands of cybersecurity.\\

\section{Methodology}
\label{sec:methodology}
This section outlines the methodology behind the development of XGen-Q, a novel large language model framework specialized for malware behavior detection and analysis in software security. XGen-Q uses the Qwen-Coder architecture and is trained with malware samples from assembly and source code. It combines RAG with a multi-step prompt approach to provide accurate classification and clear forensic explanations. Figure \ref{fig:XGen-Q} illustrates a simple interaction between input and our proposed model. The following subsections describe the dataset preparation, domain-specific pretraining, model architecture and training, prompt engineering, and semantic post-processing methods that contribute to XGen-Q’s effectiveness. Figure \ref{fig:diagram-of-framework} provides an overview of the research model.\\\\\\\\\\

\subsection{Phase 1: Dataset Collection and Preparation}

In this study, we used SBAN \cite{jelodar2025sban} dataset. The dataset contain diverse samples across multiple programming languages and malware families, allowing the model to generalize across real-world threats (see Table \ref{tab:M-datasets-size}). We utilize this dataset for pre-training the LLM. Additional details will follow in Phase 2.

\begin{table}[h]
    \centering
     \caption{Size of malware datasets (source code and assembly only) from PE files - source code (Src) and assembly (Assem). }
    \resizebox{\linewidth}{!}{
    \begin{tabular}{l c c c c}
    \hline
       Dataset  &  Source & NLD & Assem & Binary\\
       \hline
       1. BODMAS & 93711 & 93711 & 92317 & 88605\\
       2. MalwareBazzar & 14746 & 14746 & 14051 & 13973\\
       3. Sorel20m & 81584 & 81584 & 81177 & 79166\\
       4. Dike & 17431 & 17431 & 12138 & 11726\\
       5. xLangKode & 468679 & 468679 & 5974 & 13299\\
       \hline
       Total & 676151 & 676151 & 205657 & 206769\\
       \hline
    \end{tabular}}
    \label{tab:M-datasets-size}
\end{table}

\subsection{Phase 2: Domain-Specific Pre-Training}
To effectively detect and analyze malware, models need specialized training on security-focused data. Domain-specific pre-training helps the model to learn malware behaviors and threat patterns that general programming data misses. The next subsections cover why domain adaptation matters and how XGen-Q was pretrained on malware samples.

\subsubsection{Importance of Domain Adaptation}

Generic LLMs which are trained on broad programming corpora lack specialized knowledge of malware behaviors and threat signatures. Domain adaptation through targeted pretraining on security datasets enables the model to internalize malware-specific tactics, obfuscation strategies, and exploitation methods. Such adaptation improves the model’s understanding of subtle behavioral indicators that is missed by general-purpose models. It also enhances the model’s ability to explain decisions using the correct cybersecurity terminology, which is critical for trust and adoption in real-world analysis settings.

\subsubsection{Pre-training on Malware samples}

To specialize \textbf{XGen-Q}, we conduct domain-adaptive pretraining using malware samples from Table \ref{tab:M-datasets-size}. We use Causal Language Modeling (CLM) \cite{wu2024causality, zhu2023causal} for training the model. Exposure to real-world malware patterns enables the model to develop contextual awareness that supports accurate detection and interpretation of malicious behavior embedded in malware code analysis.

\subsection{Phase 3: Model Architecture and Training}

XGen-Q is built upon the \textbf{Qwen-Coder} language model, specifically the \texttt{Qwen2.5-Coder-1.5B-Instruct} variant. This version balances performance and computational efficiency, offering a 1.5B parameter model optimized for code understanding with a large token context window. The pretraining process is based on the malware samples .

\subsection{Phase 4: LLM-Prompt Design for Security Tasks}
\begin{figure*}[h]
\centering
\includegraphics[width=\linewidth]{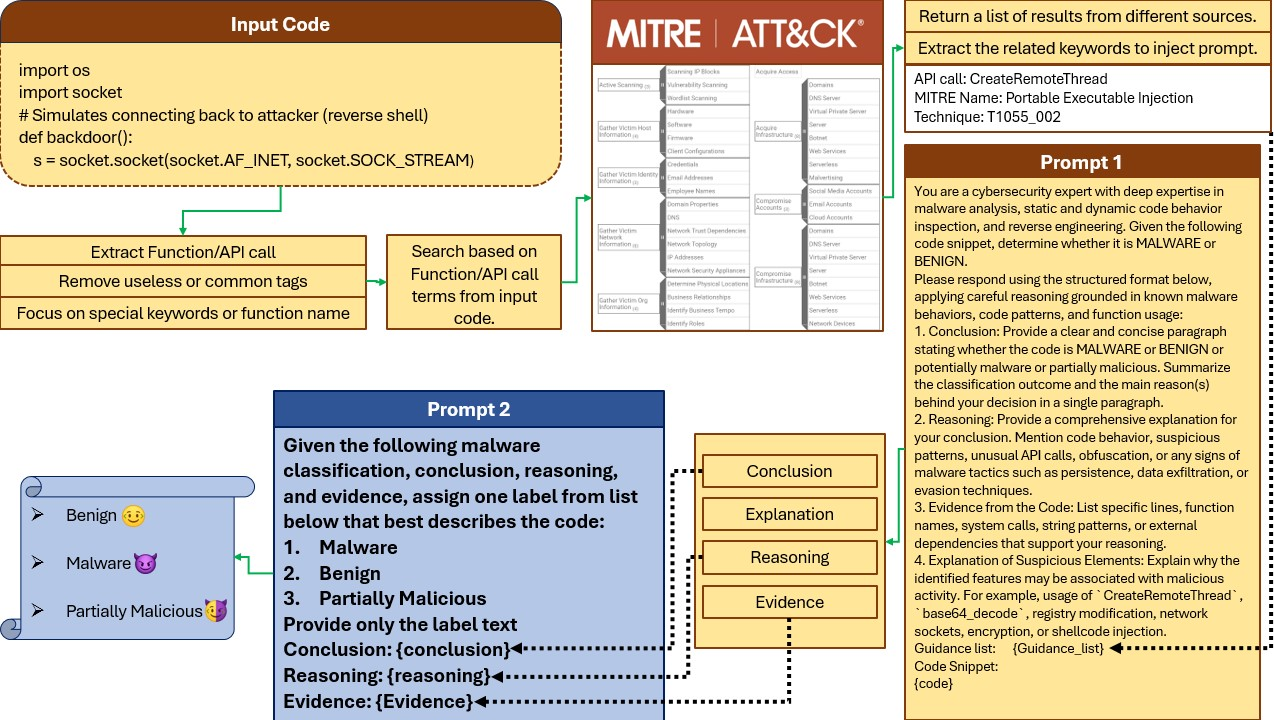}
\caption{Knowledge injection based on MITRE ATT\&CK, to build guidance for Prompt 1, and prompt 2 creation based on prompt 1's output.}
\label{fig:diagram-of-framework-example}
\end{figure*}

This model employs a two-stage prompt strategy to combine explainability with precision. Prompt 1 instructs the LLM to generate a structured forensic analysis that includes conclusion, reasoning, evidence, and suspicious behavior explanation. Prompt 2 compresses this reasoning into a single actionable label: malware, benign, or partially malicious. This approach allows flexible pipeline design, where analysts can interpret full reasoning or rely solely on classification outputs, depending on operational needs. Figure \ref{fig:diagram-of-framework-example} illustrates how an input code goes through the process.

\subsubsection{Prompt 1 - Expert Malware Behavior Classification}

Prompt 1 is structured to simulate a human cybersecurity expert’s thought process in static malware analysis. The model outputs:

\begin{enumerate}[label=\arabic*.]
    \item \textbf{Conclusion}: High-level decision on sample classification.
    \item \textbf{Reasoning}: Justification based on observed behavior.
    \item \textbf{Evidence}: Key code features such as suspicious function calls or API usage.
    \item \textbf{Explanation of Suspicious Elements}: Description of why these features are concerning.
\end{enumerate}
 
\textbf{RAG-Based Behavior Extraction (Algorithm 1)} : To enrich Prompt 1, we employ a RAG mechanism described in Algorithm 1. This algorithm extracts the top 10 behavior-relevant keywords from external knowledge  from https://attack.mitre.org/, ensuring the model is guided by the most contextually relevant security insights.

\begin{algorithm}[H]
\caption{RAG-Based Behavior Keyword Extraction.}
\begin{algorithmic}[1]
\Require Code snippet $C$
\Ensure Top 10 behavior-related keywords $K$
\State Encode $C$ into semantic vector representation $E$
\State Query a pre-indexed knowledge base (attack.mitre.org) using $E$
\State Retrieve top $N$ relevant documents $D = \{d_1, d_2, \dots, d_N\}$
\State Extract keyword candidates from $D$ using TF-IDF 
\State Rank all extracted keywords by semantic similarity to $C$
\State Select the top 10 most relevant keywords as $K$
\State \Return $K$
\end{algorithmic}
\end{algorithm}

\subsubsection{Prompt 2 - Final Behavior Classification}

Prompt 2 simplifies the structured forensic report by requesting a single classification label: malware, benign, or partially malicious. This helps to facilitate automated integration with real-time pipelines, alert systems, and triage dashboards. By separating analysis and labeling, the system allows the explanation module and decision module to evolve independently, improving flexibility for future updates or policy adjustments.\\\\

\textbf{Multi-Stage Malware Analysis Pipeline (Algorithm 2)}: Algorithm 2 outlines the complete multi-stage inference pipeline of XGen-Q, combining RAG-based retrieval, multi-step prompt generation, semantic parsing, and optional analyst review. This pipeline ensures robust, explainable classification with feedback loops for continual improvement.

\begin{algorithm}[H]
\caption{XGen-Q Multi-Stage Malware Analysis Pipeline.}
\begin{algorithmic}[1]
\Require Static code snippet $C$
\Ensure Final classification label $L \in \{$malware, benign, or partially malicious.$\}$

\State Encode $C$ into semantic vector representation $E$
\State Query a cybersecurity knowledge index using $E$
\State Retrieve top $N$ relevant documents $D$
\State Extract behavior-related keywords using NLP-based ranking
\State Select top 10 keywords $K$ for prompt injection

\State Create Prompt 1 using $C$ and $K$
\State Generate detailed forensic report $R$ from LLM
\State Parse $R$ into structured outputs:
    \begin{itemize}
        \item \texttt{Conclusion}
        \item \texttt{Reasoning}
        \item \texttt{Code Evidence}
        \item \texttt{Suspicious Element Explanation}
    \end{itemize}

\State Construct Prompt 2 using parsed output from $R$
\State Query LLM with Prompt 2 to obtain label $L$
\State Verify $L \in \{$malware, benign, or partially malicious$\}$

\State Extract and normalize fields for ingestion into SIEM/log systems
\State Store $(C, L, R)$ in a structured database
\State Use results for trend analysis and threat correlation

\If{Analyst review is enabled}
    \State Present structured report for human validation
    \State Analyst accepts/modifies classification $L$
    \State Feedback is stored for pre-training 
\EndIf

\If{Feedback available}
    \State Append corrected samples to feedback set
    \State Periodically fine-tune XGen-Q on new feedback
\EndIf

\State \Return Final classification label $L$
\end{algorithmic}
\end{algorithm}

\subsection{Phase 5: Semantic Handling and Post-Processing}

As part of the final stage in the XGen-Q pipeline, semantic handling and post-processing play a critical role in transforming unstructured LLM outputs into actionable, machine-readable formats. 

\subsubsection{Semantic Extraction}

The LLM-generated free-text output is parsed into structured fields: conclusion, reasoning, code evidence, and suspicious indicators. This parsing enables automated ingestion into monitoring systems and supports downstream analytics such as pattern tracking and correlation of threat events.

\subsubsection{Analyst Review and Auditability}

Structured outputs simplify manual review, accelerating validation and improving trust in AI-generated results. Separation between reasoning and final decision supports auditability, compliance, and post-incident investigation. Analyst's feedback is looped into future model updates, continuously improving performance and reducing false outcomes.

\section{Experiments and Implementation}
In this section, we firstly explain the configurations associated with the training our model. Then, we demonstrate the performance of our proposed model and illustrate an output generated by our framework. Finally, we compare our proposed model with similar competitors and summarize the results.
\label{sec:experiments-and-implementation}
\subsection{Settings and Computing Configuration}
All experiments were conducted on a server equipped with NVIDIA H100 GPU, offering the computational power required for large-scale model training. To optimize efficiency and reduce memory usage, we employed mixed-precision training \cite{micikevicius2017mixed,das2018mixed} during the pretraining phase.  In this study, we used LangChain \cite{langchain} to orchestrate large language model tasks and integrated \texttt{llamaindex} for implementing the framework and RAG mechanism, utilizing 'attack.mitre.org' as the primary retrieval source.

\subsection{The Performance of the Pre-training}
\begin{figure}[h]
    \centering
    \includegraphics[width=\linewidth]{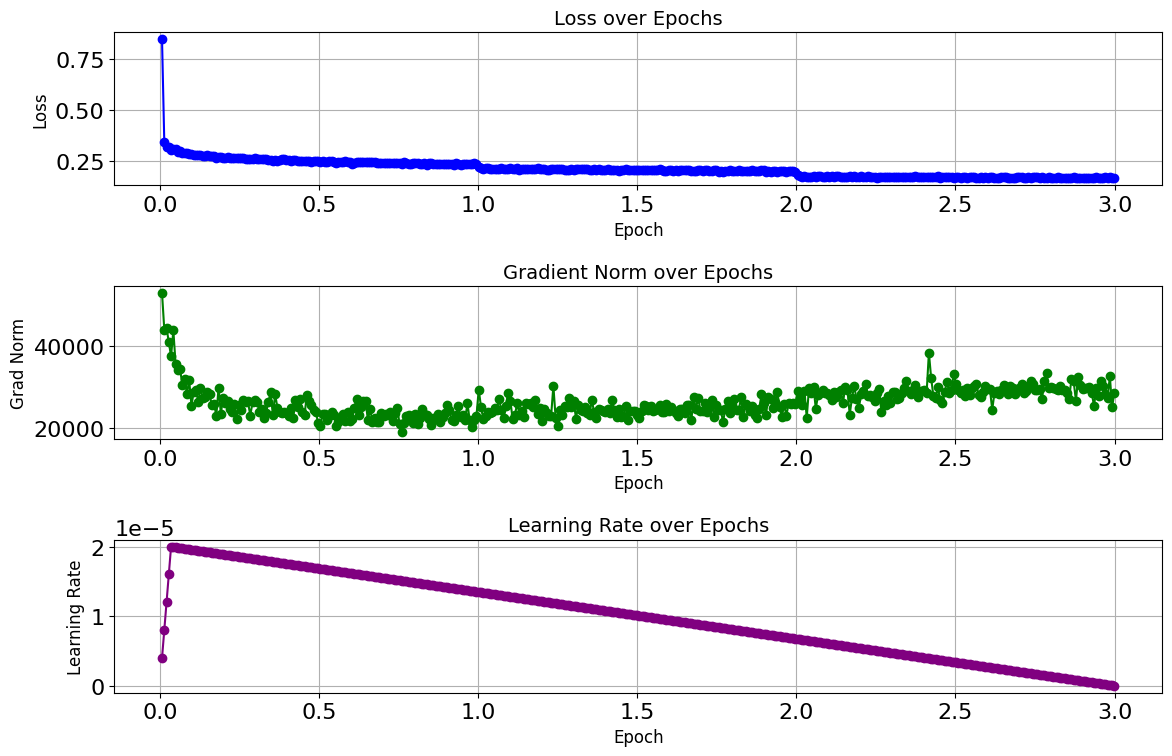}
    \caption{Training metrics of XGen-Q showing loss reduction, gradient behavior and, learning rate decay.}
    \label{fig:training-metrics}
\end{figure}

 Figure \ref{fig:training-metrics} presents three subplots that collectively illustrate the model’s training dynamics over 3 epochs. The first subplot shows a sharp drop in training loss from an initial value above 0.75 within the first 0.2 epochs, indicating rapid adaptation and effective optimization. As training progresses, the loss steadily decreases and flattens, suggesting that the model is approaching convergence with stable and robust learning behavior. The second subplot, depicting the gradient norm, begins with high values, typical for early training on new data, but quickly stabilizes, reflecting a smooth transition from large corrective updates to fine-tuning, which is a sign of healthy training. The third subplot illustrates the learning rate schedule, where an initial warmup phase allows the model to begin learning gradually, followed by a linear decay that fine-tunes updates as convergence is approached. Together, these metrics demonstrate a well-controlled and effective training process.

\subsection{Example of the XGen-Q Output}

XGen-Q generates detailed reports analyzing software samples for malicious behavior. Each report is structured to provide a clear and concise understanding of the sample's nature by breaking the analysis into four key components: conclusion, reasoning, evidence, and final Judgment as showed in Figures \ref{fig:Example-b-m} and \ref{fig:Example-pm}. This structured output facilitates both automated processing and human review, enabling security analysts to quickly assess the characteristics of a given sample.

\begin{itemize}
    \item \textbf{Conclusion} summarizes the overall assessment of the analyzed sample, directly stating whether the code appears benign, malicious, or partially malicious. This summary results from a comprehensive examination of the code’s behavior and characteristics.
    \item \textbf{Reasoning} outlines the rationale behind the conclusion, describing the analytical process and highlighting specific behaviors or patterns that led to the classification. This section promotes transparency and helps users understand the logic behind the model’s judgment.
    \item \textbf{Evidence} lists concrete findings that support the reasoning, such as suspicious API calls, code structures, or detected obfuscation techniques.
    \item \textbf{Final Judgment} provides a definitive label, categorizing the sample as malware, benign, or partially malicious. This final decision streamlines response workflows by offering a clear and actionable verdict.
\end{itemize}

\begin{figure}[!htbp]


\begin{lstlisting}[language=json,caption={ }]
{
  "ID": "malware_sample_0645470.c",
  "conclusion": "Classified as MALWARE.",
  "reasoning": "Suspicious use of Windows Update... ",
  "evidence": [
    "CreateProcessA used to execute update.exe.",
  ],
  "final_Judgment": "MALWARE",
  "source_code": "oid exploitWindowsUpdate() { ... }\nint main() { exploitWindowsUpdate(); return 0; }"
}
\end{lstlisting}

\caption{Example of a sample considered as malware.}
\label{fig:Example-b-m}
\end{figure}

\begin{figure}

\begin{lstlisting}[language=json,caption={}]
{
  "ID": "malware_sample_0816286.c",
  "conclusion": "This code is neither clearly MALWARE nor BENIGN. ",
  "reasoning": "The use of encrypted DLL injection via CreateRemoteThread suggests evasive behavior.",
  "evidence": [
    "Encrypted DLL loaded using LoadLibraryA.",
  ],
  "final_Judgment": "PARTIALLY MALICIOUS",
  "source_code": "void inject_polymorphic_dll(DWORD pid) { ... }\nint main() { ... }"
}
\end{lstlisting}
\caption{Example of a sample considered as partially malicious.}
\label{fig:Example-pm}
\end{figure}

\subsection{Comparing XGen-Q with Existing Code Models}

To demonstrate how well the model understands malware code, we use the \emph{perplexity} metric. We evaluated several LLMs based on this metric to compare their performance. For ease of reference, the models \textbf{LM-3}, \textbf{DS-1.3B}, and \textbf{Phi-4} correspond respectively to the full model identifiers \texttt{meta-llama/Llama-3.1-8B-Instruct}~\cite{touvron2023llama}, \texttt{deepseek-ai/deepseek-coder-\\1.3b-instruct}~\cite{liu2024deepseek}~\cite{abouelenin2025phi}, and \texttt{microsoft/Phi-4-mini-instruct}.\\

\begin{figure}[h]
    \centering
    \includegraphics[width=\linewidth]{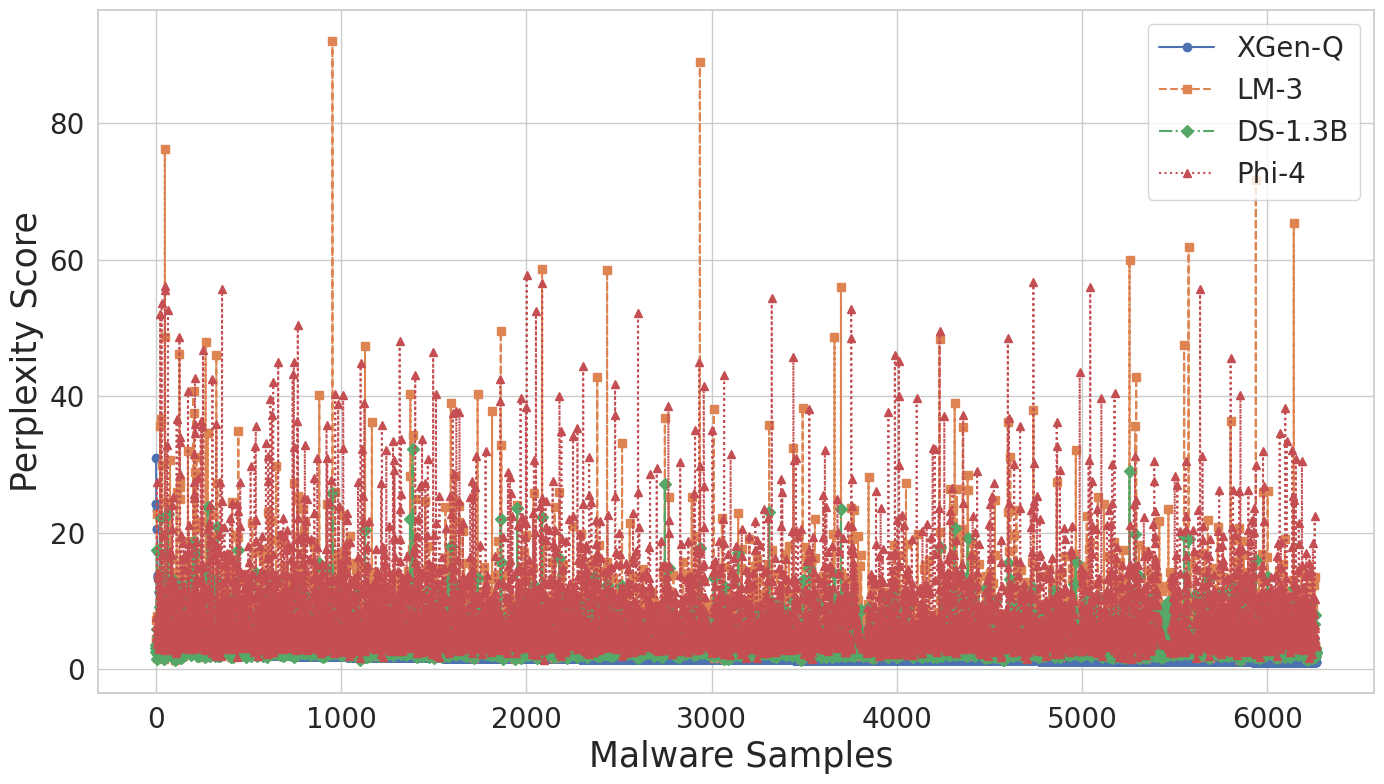}
    \caption{Perplexity values from four models on malware source code samples.}
    \label{fig:erplexit}
\end{figure}

As shown in Figure~\ref{fig:erplexit}, XGen-Q demonstrates superior performance, consistently maintaining lower perplexity scores across all sample sizes compared to LM-3, DS-1.3B, and Phi-4. Its curve shows a steep initial decline and stabilizes at an impressively low level, indicating excellent generalization capability even as the dataset scales to 6,000 samples. This robust performance suggests that XGen-Q is particularly well-optimized for handling large-scale, complex tasks efficiently. While LM-3 also performs respectably, its perplexity scores remain slightly higher than XGen-Q’s, especially at larger sample sizes. The gap between these two models widens as data volume increases, reinforcing XGen-Q’s scalability advantage. For applications requiring both precision and the ability to process massive datasets, XGen-Q clearly emerges as the top choice, its combination of low perplexity and stability makes it the standout model in this comparison.

\begin{figure}[h]
    \centering
    \includegraphics[width=\linewidth]{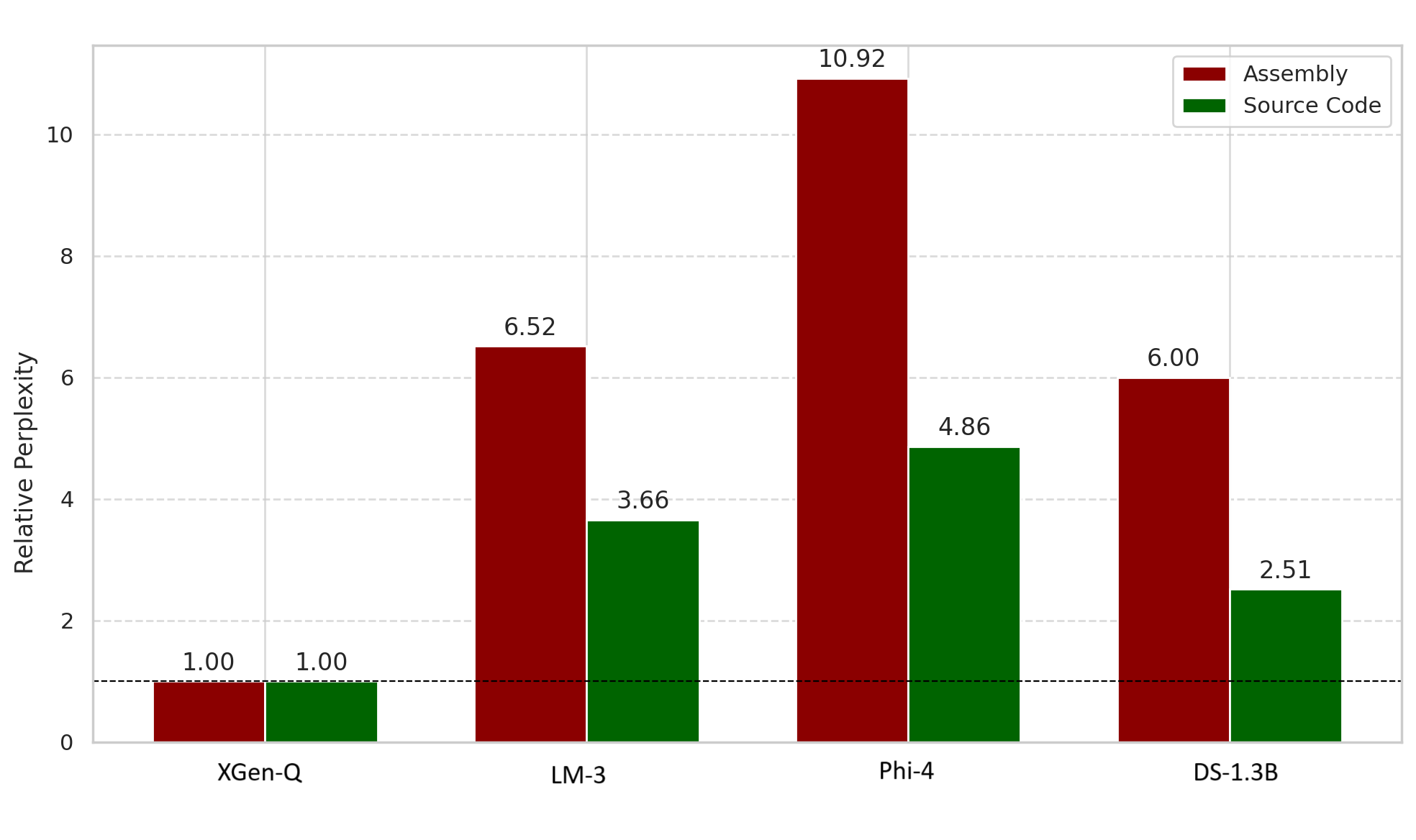}
    \caption{Line graph with distinct markers showing individual perplexity scores.}
    \label{fig:stack}
\end{figure}



\begin{table}[ht]
\centering
\caption{Perplexity comparison of language models on assembly and source code levels. Lower perplexity indicates better performance.}
\resizebox{\linewidth}{!}{%
\begin{tabular}{l l c c}
\toprule
\textbf{Data} & \textbf{Model} & \textbf{Perplexity} $\downarrow$ & \textbf{Relative to XGen-Q (×)} \\
\midrule
\multirow{4}{*}{Assembly} 
  & \textbf{XGen-Q} & \textbf{1.530} & \textbf{1.00× (best)} \\
  & LM-3                    & 9.972          & 6.52× \\
  & Phi-4                  & 16.713         & 10.93× \\
  & DS-1.3B                & 9.183          & 6.00× \\
\midrule
\multirow{4}{*}{Source Code} 
  & \textbf{XGen-Q} & \textbf{1.592} & \textbf{1.00× (best)} \\
  & LM-3                    & 5.822          & 3.66× \\
  & Phi-4                  & 7.739          & 4.86× \\
  & DS-1.3B                & 3.997          & 2.51× \\
\bottomrule
\end{tabular}
}
\label{tab:perplexity_comparison}
\end{table}

Figure \ref{fig:stack} and Table \ref{tab:perplexity_comparison} summarize the comparative performance of all evaluated models on both assembly and source code variants of the same malware samples. Across both code types, \textbf{XGen-Q} achieves the lowest perplexity scores, outperforming all baselines. On assembly code, XGen-Q scores 1.530, substantially better than \textbf{DS-1.3B} (9.183) and \textbf{Phi-4} (16.713). On source code, it again leads with 1.592, while the next-best baseline (\textbf{DS-1.3B}) trails at 3.997. These results clearly demonstrate \textbf{XGen-Q}'s domain-specific strength in modeling malware code, which is often obfuscated, irregular, and semantically complex.

\section{Discussion, Limitations and Future Works}
\label{sec:limitations-and-future-works}
The XGen-Q framework demonstrates the feasibility and benefits of combining domain-specific pre-training, RAG, and multi-stage prompt engineering for enhanced software security analysis. Our evaluation highlights that adapting a strong base LLM like Qwen, using carefully curated malware and benign code datasets, leads to substantial improvements in detection, accuracy, and interpretability. The integration of RAG allows the model to remain current with the evolving threat landscape by dynamically incorporating external knowledge at inference time.\\ We also mentioned that in this work, we limited the size of the dataset used for pre-training the LLM. Future work will explore incorporating multi-modal inputs, such as dynamic execution traces, binary metadata, and network telemetry, to further enrich the model’s contextual understanding \cite{gebrehans2025generative,khorrami2025real, shafi2024intruders}. We also aim to investigate continual learning strategies that allow XGen-Q to incrementally update its internal knowledge without requiring full retraining. Moreover, integrating explainable AI techniques will improve the transparency of the model’s predictions, making its outputs more actionable and trustworthy for security analysts \cite{jelodar2025large,jelodar2025largesource}.

\section{Conclusion}
\label{sec:conclusion}
This paper presents XGen-Q, a domain-adapted large language model built on the Qwen-Coder architecture for robust and interpretable malware analysis. Trained on a large-scale dataset of over one million malware samples and enhanced through domain-specific pretraining techniques, XGen-Q is capable of capturing complex and subtle code patterns associated with malicious behavior. By incorporating retrieval-augmented generation (RAG) and a multi-stage inference strategy, the model supports detailed and context-aware forensic reporting, even in the presence of advanced code obfuscation. Experimental results demonstrate that XGen-Q achieves lower perplexity and strong generalization to previously unseen samples. These findings underscore the potential of XGen-Q as a powerful and explainable tool for advancing automated malware analysis.


\bibliographystyle{ieeetr}
\bibliography{main}

\begin{thebibliography}{10}

\bibitem{bohme2025software}
M.~B{\"o}hme and E.~Bodden, ``Software security analysis in 2030 and beyond: A research roadmap,'' {\em ACM Transactions on Software Engineering and Methodology}, vol.~34, no.~5, pp.~1--26, 2025.

\bibitem{rahimi2025comprehensive}
N.~Rahimi, B.-A. Schuelke-Leech, and M.~Mirhassani, ``A comprehensive review of security vulnerabilities in heavy-duty vehicles: Comparative insights and current research gaps,'' {\em Computers \& Security}, p.~104452, 2025.

\bibitem{mohamed2025artificial}
N.~Mohamed, ``Artificial intelligence and machine learning in cybersecurity: a deep dive into state-of-the-art techniques and future paradigms,'' {\em Knowledge and Information Systems}, pp.~1--87, 2025.

\bibitem{aldasoro2025generative}
I.~Aldasoro, S.~Doerr, L.~Gambacorta, S.~Notra, T.~Oliviero, and D.~Whyte, ``Generative artificial intelligence and cyber security in central banking,'' {\em Journal of Financial Regulation}, vol.~11, no.~1, pp.~119--128, 2025.

\bibitem{rahman2025cyber}
M.~M. Rahman, S.~Hossain, B.~Bhusal, and N.~Kshetri, ``Cyber ai trends: Future trends in ai for cyberbullying prevention,'' in {\em Combating Cyberbullying With Generative AI}, pp.~279--298, IGI Global Scientific Publishing, 2025.

\bibitem{xu2024large}
H.~Xu, S.~Wang, N.~Li, K.~Wang, Y.~Zhao, K.~Chen, T.~Yu, Y.~Liu, and H.~Wang, ``Large language models for cyber security: A systematic literature review,'' {\em arXiv preprint arXiv:2405.04760}, 2024.

\bibitem{tian2025exploring}
S.~Tian, T.~Zhang, J.~Liu, J.~Wang, X.~Wu, X.~Zhu, R.~Zhang, W.~Zhang, Z.~Yuan, S.~Mao, {\em et~al.}, ``Exploring the role of large language models in cybersecurity: A systematic survey,'' {\em arXiv preprint arXiv:2504.15622}, 2025.

\bibitem{kasri2025vulnerability}
W.~Kasri, Y.~Himeur, H.~A. Alkhazaleh, S.~Tarapiah, S.~Atalla, W.~Mansoor, and H.~Al-Ahmad, ``From vulnerability to defense: The role of large language models in enhancing cybersecurity,'' {\em Computation}, vol.~13, no.~2, p.~30, 2025.

\bibitem{hossain2024malicious}
A.~A. Hossain, M.~K. PK, J.~Zhang, and F.~Amsaad, ``Malicious code detection using llm,'' in {\em NAECON 2024-IEEE National Aerospace and Electronics Conference}, pp.~414--416, IEEE, 2024.

\bibitem{al2024exploring}
J.~Al-Karaki, M.~A.-Z. Khan, and M.~Omar, ``Exploring llms for malware detection: Review, framework design, and countermeasure approaches,'' {\em arXiv preprint arXiv:2409.07587}, 2024.

\bibitem{zhou2025srdc}
C.~Zhou, Y.~Liu, W.~Meng, S.~Tao, W.~Tian, F.~Yao, X.~Li, T.~Han, B.~Chen, and H.~Yang, ``Srdc: Semantics-based ransomware detection and classification with llm-assisted pre-training,'' in {\em Proceedings of the AAAI Conference on Artificial Intelligence}, vol.~39, pp.~28566--28574, 2025.

\bibitem{qian2025lamd}
X.~Qian, X.~Zheng, Y.~He, S.~Yang, and L.~Cavallaro, ``Lamd: Context-driven android malware detection and classification with llms,'' {\em arXiv preprint arXiv:2502.13055}, 2025.

\bibitem{yu2024maltracker}
Z.~Yu, M.~Wen, X.~Guo, and H.~Jin, ``Maltracker: A fine-grained npm malware tracker copiloted by llm-enhanced dataset,'' in {\em Proceedings of the 33rd ACM SIGSOFT International Symposium on Software Testing and Analysis}, pp.~1759--1771, 2024.

\bibitem{hasanov2024application}
I.~Hasanov, S.~Virtanen, A.~Hakkala, and J.~Isoaho, ``Application of large language models in cybersecurity: A systematic literature review,'' {\em IEEE Access}, 2024.

\bibitem{zhang2025llms}
J.~Zhang, H.~Bu, H.~Wen, Y.~Liu, H.~Fei, R.~Xi, L.~Li, Y.~Yang, H.~Zhu, and D.~Meng, ``When llms meet cybersecurity: A systematic literature review,'' {\em Cybersecurity}, vol.~8, no.~1, pp.~1--41, 2025.

\bibitem{feng2025llm}
R.~Feng, H.~Chen, S.~Wang, M.~M. Karim, and Q.~Jiang, ``Llm-maldetect: A large language model-based method for android malware detection,'' {\em IEEE Access}, 2025.

\bibitem{lee2025trag}
D.~Lee, J.~Kim, J.~Kim, S.-w. Hwang, and J.~Park, ``trag: Term-level retrieval-augmented generation for domain-adaptive retrieval,'' in {\em Proceedings of the 2025 Conference of the Nations of the Americas Chapter of the Association for Computational Linguistics: Human Language Technologies (Volume 1: Long Papers)}, pp.~6566--6578, 2025.

\bibitem{long2023adapt}
Q.~Long, W.~Wang, and S.~J. Pan, ``Adapt in contexts: Retrieval-augmented domain adaptation via in-context learning,'' {\em arXiv preprint arXiv:2311.11551}, 2023.

\bibitem{leemann2024auto}
T.~Leemann, P.~Petridis, G.~Vietri, D.~Manousakas, A.~Roth, and S.~Aydore, ``Auto-gda: Automatic domain adaptation for efficient grounding verification in retrieval augmented generation,'' {\em arXiv preprint arXiv:2410.03461}, 2024.

\bibitem{xu2024simrag}
R.~Xu, H.~Liu, S.~Nag, Z.~Dai, Y.~Xie, X.~Tang, C.~Luo, Y.~Li, J.~C. Ho, C.~Yang, {\em et~al.}, ``Simrag: Self-improving retrieval-augmented generation for adapting large language models to specialized domains,'' {\em arXiv preprint arXiv:2410.17952}, 2024.

\bibitem{lewis2020retrieval}
P.~Lewis, E.~Perez, A.~Piktus, F.~Petroni, V.~Karpukhin, N.~Goyal, H.~K{\"u}ttler, M.~Lewis, W.-t. Yih, T.~Rockt{\"a}schel, {\em et~al.}, ``Retrieval-augmented generation for knowledge-intensive nlp tasks,'' {\em Advances in neural information processing systems}, vol.~33, pp.~9459--9474, 2020.

\bibitem{brown2020language}
T.~Brown, Mann, {\em et~al.}, ``Language models are few-shot learners,'' {\em Advances in neural information processing systems}, vol.~33, pp.~1877--1901, 2020.

\bibitem{xu2024detecting}
Z.~Xu and V.~S. Sheng, ``Detecting ai-generated code assignments using perplexity of large language models,'' in {\em Proceedings of the aaai conference on artificial intelligence}, vol.~38, pp.~23155--23162, 2024.

\bibitem{cooper2024perplexed}
N.~Cooper and T.~Scholak, ``Perplexed: Understanding when large language models are confused,'' {\em arXiv preprint arXiv:2404.06634}, 2024.

\bibitem{xu2024investigating}
J.~Xu, H.~Zhang, Y.~Yang, Z.~Cheng, J.~Lyu, B.~Liu, X.~Zhou, L.~Yang, A.~Bacchelli, Y.~K. Chiam, {\em et~al.}, ``Investigating efficacy of perplexity in detecting llm-generated code,'' {\em arXiv preprint arXiv:2412.16525}, 2024.

\bibitem{yusof2024code}
M.~A. Yusof and S.~Saee, ``Code switching: exploring perplexity and coherence metrics for optimizing topic models of historical documents,'' {\em International Journal of Systematic Innovation}, vol.~8, no.~4, pp.~103--118, 2024.

\bibitem{jelodar2025sban}
H.~Jelodar, M.~Meymani, S.~Bai, R.~Razavi-Far, and A.~A. Ghorbani, ``Sban: A framework \& multi-dimensional dataset for large language model pre-training and software code mining,'' in {\em Proceedings of the 2025 IEEE International Conference on Data Mining Workshops (ICDMW)}, IEEE, 2025.

\bibitem{wu2024causality}
A.~Wu, K.~Kuang, M.~Zhu, Y.~Wang, Y.~Zheng, K.~Han, B.~Li, G.~Chen, F.~Wu, and K.~Zhang, ``Causality for large language models,'' {\em arXiv preprint arXiv:2410.15319}, 2024.

\bibitem{zhu2023causal}
Z.~Zhu, H.~Yu, C.~Shen, J.~Du, Z.~Shen, and Z.~Wang, ``Causal language model aided sequential decoding with natural redundancy,'' {\em IEEE Transactions on Communications}, vol.~71, no.~5, pp.~2685--2697, 2023.

\bibitem{micikevicius2017mixed}
P.~Micikevicius, S.~Narang, J.~Alben, G.~Diamos, E.~Elsen, D.~Garcia, B.~Ginsburg, M.~Houston, O.~Kuchaiev, G.~Venkatesh, {\em et~al.}, ``Mixed precision training,'' {\em arXiv preprint arXiv:1710.03740}, 2017.

\bibitem{das2018mixed}
D.~Das, N.~Mellempudi, D.~Mudigere, D.~Kalamkar, S.~Avancha, K.~Banerjee, S.~Sridharan, K.~Vaidyanathan, B.~Kaul, E.~Georganas, {\em et~al.}, ``Mixed precision training of convolutional neural networks using integer operations,'' {\em arXiv preprint arXiv:1802.00930}, 2018.

\bibitem{langchain}
H.~Chase and L.~Contributors, ``Langchain,'' 2022.

\bibitem{touvron2023llama}
H.~Touvron, T.~Lavril, G.~Izacard, X.~Martinet, M.-A. Lachaux, T.~Lacroix, B.~Rozi{\`e}re, N.~Goyal, E.~Hambro, F.~Azhar, {\em et~al.}, ``Llama: Open and efficient foundation language models,'' {\em arXiv preprint arXiv:2302.13971}, 2023.

\bibitem{liu2024deepseek}
A.~Liu, B.~Feng, B.~Xue, B.~Wang, B.~Wu, C.~Lu, C.~Zhao, C.~Deng, C.~Zhang, C.~Ruan, {\em et~al.}, ``Deepseek-v3 technical report,'' {\em arXiv preprint arXiv:2412.19437}, 2024.

\bibitem{abouelenin2025phi}
A.~Abouelenin, A.~Ashfaq, A.~Atkinson, H.~Awadalla, N.~Bach, J.~Bao, A.~Benhaim, M.~Cai, V.~Chaudhary, C.~Chen, {\em et~al.}, ``Phi-4-mini technical report: Compact yet powerful multimodal language models via mixture-of-loras,'' {\em arXiv preprint arXiv:2503.01743}, 2025.

\bibitem{gebrehans2025generative}
G.~Gebrehans, N.~Ilyas, K.~Eledlebi, W.~T. Lunardi, M.~Andreoni, C.~Y. Yeun, and E.~Damiani, ``Generative adversarial networks for dynamic malware behavior: A comprehensive review, categorization, and analysis,'' {\em IEEE Transactions on Artificial Intelligence}, 2025.

\bibitem{khorrami2025real}
F.~Khorrami, R.~Karri, and P.~Krishnamurthy, ``Real-time multi-modal subcomponent-level measurements for trustworthy system monitoring and malware detection,'' {\em arXiv preprint arXiv:2501.13081}, 2025.

\bibitem{shafi2024intruders}
M.~Shafi, ``Intruders' behavior unveiled: A dual-tier behavior-driven model for malicious activity detection in iot network using graph learning,'' 2024.

\bibitem{jelodar2025large}
H.~Jelodar, S.~Bai, P.~Hamedi, H.~Mohammadian, R.~Razavi-Far, and A.~Ghorbani, ``Large language model (llm) for software security: Code analysis, malware analysis, reverse engineering,'' {\em arXiv preprint arXiv:2504.07137}, 2025.

\bibitem{jelodar2025largesource}
H.~Jelodar, M.~Meymani, and R.~Razavi-Far, ``Large language models (llms) for source code analysis: applications, models and datasets,'' {\em arXiv preprint arXiv:2503.17502}, 2025.

\end{thebibliography}

\end{document}